\documentclass[12pt]{article}
\usepackage[centertags]{amsmath}
\usepackage{amsfonts}
\usepackage{amssymb}
\usepackage{amsthm}
\usepackage{newlfont}
\usepackage{epsfig}
\usepackage{amscd}

\theoremstyle{plain}
\newtheorem{Th}{Theorem}
\newtheorem{Cor}[Th]{Corollary}

\newtheorem{Prop}[Th]{Proposition}
\theoremstyle{definition}
\newtheorem{Def}{Definition}

\theoremstyle{remark}
\newtheorem*{Rem}{Remark}
\newcommand{\PP}{{\mathbb P}}
\newcommand{\RR}{{\mathbb R}}
\newcommand{\EE}{{\mathbb E}}
\newcommand{\ZZ}{{\mathbb Z}}

\newcommand{\cL}{{\mathcal L}}
\newcommand{\cQ}{{\mathcal Q}}

\newcommand{\bx}{{\boldsymbol x}}
\newcommand{\boe}{{\boldsymbol e}}
\newcommand{\bN}{{\boldsymbol N}}
\newcommand{\by}{{\boldsymbol y}}
\newcommand{\bu}{{\boldsymbol u}}
\newcommand{\bv}{{\boldsymbol v}}

\newcommand{\gp}{{\mathfrak p}}
\newcommand{\gq}{{\mathfrak q}}

\newcommand{\tgp}{{\tilde\gp}}
\newcommand{\ta}{\tilde{a}}
\newcommand{\tb}{\tilde{b}}

\newcommand{\tA}{\tilde{A}}
\newcommand{\tB}{\tilde{B}}
\newcommand{\tC}{\tilde{C}}

\newcommand{\D}{{\Delta}}
\newcommand{\tD}{\tilde{\Delta}}
\newcommand{\pp}{{\partial}}

\begin{document}

\title{ Discrete asymptotic nets and W-congruences
in Pl\"{u}cker line geometry}

\author{Adam Doliwa\thanks{Supported in part by the Polish Committee 
of Scientific
Research (KBN) under Grant No. 2-P03B-143-15} \\ 
{\it Instytut Fizyki Teoretycznej, Uniwersytet Warszawski}\\
{\it ul. Ho\.{z}a 69, 00-681 Warszawa, Poland}\\ 
e-mail: {\tt doliwa@fuw.edu.pl}}
\date{}
\maketitle

\begin{abstract}

\noindent The asymptotic lattices and their transformations
are studied within the line geometry approach.
It is shown that the discrete asymptotic nets are represented by
isotropic congruences in the Pl\"ucker quadric. On the basis of 
the Lelieuvre-type representation of asymptotic lattices and of the 
discrete analog of the Moutard transformation, it is constructed
the discrete analog of the W--congruences, which provide the 
Darboux--B\"{a}cklund type transformation of asymptotic lattices.
The permutability theorems for the discrete Moutard transformation and for
the corresponding transformation of asymptotic lattices are
established as well.
Moreover it is proven that the discrete
W--congruences are represented by quadrilateral lattices in 
the quadric of Pl\"ucker. These results generalize to a discrete level
the classical line-geometric approach to asymptotic nets and W--congruences,
and incorporate the theory of
asymptotic lattices into more general theory of quadrilateral lattices and
their reductions. \\

{\it 1998 PACS:} 04.60.Nc, 02.40.Hw

\end{abstract}

\section{Introduction}

The modern theory of integrable
partial differential equations is closely related to the XIX century
differential geometry as presented in monographs of Bianchi and
Darboux~\cite{Bianchi,DarbouxIV}.
In that classical period many geometers studied "interesting"
classes of surfaces. A remarkable property of these 
surfaces (or more appropriate: coordinate systems on surfaces and 
submanifolds)
is that they allow for transformations, which
exhibit the so called {\it permutability property}.
Such transformations called, depending on the
context, the Darboux, Bianchi, B\"acklund, Laplace, Moutard, 
Combescure, L\'evy, Ribaucour or fundamental transformations of Jonas,
can be also described in terms of certain families of lines called line
congruences~\cite{Eisenhart-TS,Finikov}.

To give an example, the angle between
the asymptotic directions on the pseudospherical surfaces in $\EE^3$, when
written as a function of the asymptotic coordinates satisfies the
sine-Gordon equation. From this point of view the study of 
pseudospherical surfaces is,
roughly speaking, equivalent to studying of the sine-Gordon equation
and its solutions. The transformations of pseudospherical surfaces,
introduced 
by Bianchi and B\"acklund, lead to the celebrated B\"acklund transformations
of the sine-Gordon equation.

At the end of XIX-th century it was also discovered that most of
the "interesting" submanifolds are provided by
reductions of conjugate nets (see Section~\ref{sec:conj-nets}), and
the transformations between such submanifolds are the corresponding 
reductions of the fundamental (or Jonas) transformations of conjugate nets. 
It is worth of mentioning that from the point of view of integrable 
systems the conjugate nets and their iso-conjugate deformations
and transformations are described by the so called
multicomponent Kadomtsev--Petviashvilii 
hierarchy~\cite{DMMMS}. 

Apparently, asymptotic nets seem not to be 
directly related to conjugate nets. However, there exists an approach to
asymptotic nets and their transformations (W--congruences) describing them
as conjugate nets within the line geometry of Pl\"ucker; see
Sections~\ref{sec:line-geometry} and \ref{sec:ass-nets} for more details. 

In the soliton theory the
discrete integrable systems are considered more fundamental then the
corresponding differential systems \cite{SIDEI,SIDEII,SIDEIII}. 
Discrete equations include the
continuous theory as the result of a limiting procedure, moreover different
limits can give from one discrete equation variuous differential ones. 
Furthermore,
discrete equations reveal some symmetries lost in the continuous limit.

During last few years the connection between geometry and integrability
has been observed also at a discrete level. 
It turns out that the discrete analogs of pseudospherical surfaces were 
studied long time ago by Sauer; see~\cite{Sauer} and references therein. 
In connection with the Hirota discrete analog of
the sine-Gordon equation~\cite{HirotaSG} these "discrete pseudospherical 
surfaces" were
investigated by Bobenko and Pinkall~\cite{BP1}. 
In the book of Sauer~\cite{Sauer} one can find also other examples of
discrete surfaces, or better $\ZZ^2$ lattices
in $\RR^3$; in particular, he defined {\it discrete asymptotic nets} and 
{\it discrete conjugate nets} (consult also
Sections~\ref{sec:conj-nets} and~\ref{sec:DA}). These definitions, not only
have clear geometric meaning, but also provide
the proper, from the point of view of integrability, 
discretizations of asymptotic and conjugate nets on surfaces.

The importance of discrete conjugate nets in integrability theory was
recognized in~\cite{DCN}, where it was demonstrated that
(the discrete analog
of) the Laplace sequence of such lattices provides 
geometric interpretation of Hirota's discretization of the two dimensional 
Toda system~\cite{Hirota} -- one of the most important equations of 
the soliton 
theory and its applications.
Soon after that Doliwa and Santini defined and studied~\cite{MQL} the
discrete analogs of multidimensional conjugate nets (multidimensional
quadrilateral
lattices). They also found that the corresponding equations were already
known in the literature, being obtained by Bogdanov and
Konopelchenko~\cite{BoKo} from the $\bar\pp$ approach.

The Darboux-type transformations of the quadrilateral lattices have been
found by Ma\~nas, Doliwa and Santini~\cite{MDS}. The same authors also
investigated in detail the geometry of these transformations~\cite{TQL};
in order to do that 
the theory of discrete congruences has been constructed as well. 

In recent literature one can find various examples of integrable
discrete geometries (see, for example,~\cite{Dol-RC,q-red,DS-sym} and
articles in~\cite{BobenkoSeiler}). It turns out
that all the known, up to now, integrable lattices are special cases of 
asymptotic or quadrilateral lattices. 
For example, discrete pseudospherical 
surfaces
investigated by Bobenko and Pinkall~\cite{BP1} and
discrete affine spheres considered by Bobenko and 
Schief~\cite{BobenkoSchief-DAS} are asymptotic lattices
subjected to additional constraints. 

Given a physical system described by 
integrable partial differential equations, then 
one of first steps towards quantizing the model is to find its
discrete version preserving the integrability properties (see, for
example, \cite{KBI} and references therein). It turns out that often (see,
for example, discussion in \cite{KLWZ}) information coming from the quantum
model arises naturally as a result of the solution of the classical discrete
integrable equations. 

Some recent attempts to quantize the theory of gravity use approach of
fluctuating geometries (see recent reviews \cite{KauffmanSmolin,AJL}) based
on the concept of discrete manifolds. However most of the research
in this direction is done by computer simulations, therefore 
examples of lattice geometries described by integrable equations may be of
some help in developing this program. Such integrable lattice geometries 
could be then
studied using powerful tools of the soliton theory, such like the (quantum)
inverse spectral transform, algebro-geometric methods of integration, etc.

The connection of asymptotic nets with stationary axially symmetric 
solutions of the Einstein equations \cite{Ernst} is well known in the 
literature (see, for example, \cite{LeviSym}). Recently there was 
discovered by Schief \cite{Schief-SDE-Tz-2} an intriguing link between 
self-dual Einstein spaces \cite{Plebanski} and discrete affine spheres,
which form an integrable subcase of discrete asymptotic nets. It is
therefore reasonable to study integrability of general asymptotic lattices.

The main results of this paper are contained Theorems~\ref{th:D-A} and
\ref{th:D-W}, which incorporate the theory of
asymptotic lattices and their transformations into the theory of 
quadrilateral lattices. These results are direct analogs of the 
above-mentioned
approach to asymptotic nets in terms of conjugate nets in Pl\"ucker
quadric. The direct proof of integrability of asymptotic lattices, which
does not use the theory of quadrilateral lattices, is contained in
permutability Theorems \ref{th:perm-M} and \ref{th:perm-W}.

More detailed description of results is given below. 
\begin{itemize}
\item Asymptotic 
lattices are represented in Pl\"ucker
quadric by isotropic congruences. 
\item The asymptotic tangents 
are represented by focal lattices of such congruences. 
\item The Darboux-B\"acklund transformations of asymptotic lattices 
are provided by a discrete analog of W--congruences. 
\item Discrete W--congruences can 
be constructed from the discrete Moutard transformation via the discrete
analog of the Lelieuvre formulas  
introduced in~\cite{KoPin,NieszporskiDA}.
\item The discrete W--congruences are represented in the Pl\"ucker quadric 
by quadrilateral lattices.
\item The discussed transformations of asymptotic lattices satisfy the
permutability property.
\end{itemize}

To make our exposition self-contained we first recall necessary results 
of the
theory of conjugate nets and quadrilateral lattices
(Section~\ref{sec:conj-nets}) and basic notions of the line geometry
(Section~\ref{sec:line-geometry}). Section~\ref{sec:ass-nets} is intended to
motivate our investigations and contains a brief summary of the theory
of asymptotic nets and W--congruences.
In Section~\ref{sec:DA} we construct the
theory of discrete asymptotic nets within the line geometry of Pl\"ucker.
Section~\ref{sec:DW} provides a detailed exposition
of the discrete W--congruences. Finally, in Section \ref{sec:perm} we state and
prove the permutability theorems for the Moutard transformation and for the
corresponding $W$--transformation of asymptotic lattices.

\section{Quadrilateral lattices and congruences}
\label{sec:conj-nets}
In this Section we present basic result from the theory of conjugate nets
and
congruences~\cite{Eisenhart-TS,DarbouxOS,Lane}, and their discrete 
generalizations~\cite{Sauer,DCN,MQL,TQL}. We give here only 
definitions necessary to understand results of this paper. In particular, 
we consider only two dimensional conjugate nets and lattices. 

\begin{figure}
\begin{center}
\epsffile{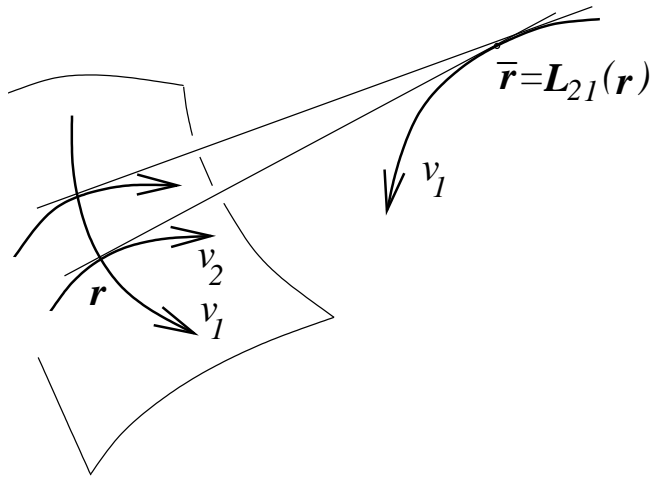}
\end{center}
\caption{Conjugate net}
\label{fig:conj}
\end{figure}

\begin{Def}
A coordinate system on a surface in $\PP^M$, is called {\it conjugate net}
if tangents to any parametric line transported in the second direction form
a developable surface (see Fig.~\ref{fig:conj}).
\end{Def} 
This geometric characterization can be put into form of the Laplace equation
satisfied by homogeneous coordinates $\by(v_1,v_2)\in\RR^{M+1}$ of the 
net
\begin{equation} \label{eq:Lap}
\pp_1\pp_2 \by = a \pp_1\by + b \pp_2\by + c \by ,
\end{equation}
here $v_1$, $v_2$ are the conjugate parameters, $\pp_i$ denotes the partial
derivative with respect to $v_i$, $i=1,2$, and $a(v_1,v_2)$, $b(v_1,v_2)$,
$c(v_1,v_2)$ are functions of the conjugate parameters.
Given conjugate net on a surface, it defines two new conjugate nets called
the Laplace transforms of the old net; the transformations are provided by 
tangents to the parametric lines, see Fig.~\ref{fig:conj}. 

\begin{figure}
\begin{center}
\epsffile{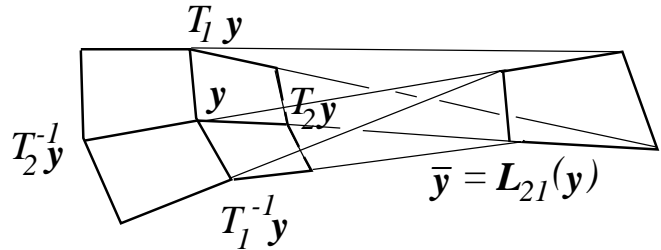}
\end{center}
\caption{Quadrilateral surface}
\label{fig:quadr2}
\end{figure}

The discrete version of conjugate net on a surface is given by two
dimensional quadrilateral lattice (quadrilateral surface).
\begin{Def}
By {\it quadrilateral surface} we mean mapping of $\ZZ^2$ in $\PP^M$, such
that its elementary quadrilaterals are planar (see Fig.~\ref{fig:quadr2}).
\end{Def} 

\begin{Rem}
Notice that tangents to any parametric discrete curve transported in the 
second direction form a discrete analog of a developable surface, i.e.,
one-parameter family of lines tangent to a (discrete) curve.
\end{Rem}
This geometric characterization implies linear relation
between homogeneous coordinates $\by(m_1,m_2)\in\RR^{M+1}$ of four points of
any elementary quadrilateral with vertices $\by$, $T_1\by$, $T_2\by$ and
$T_1 T_2 \by$, where $T_i$ denotes shift operator along $i$-th direction of
the lattice, $i=1,2$. Such a relation can be put into the form of the
discrete Laplace equation  
\begin{equation} \label{eq:D-Lap}
\D_1\D_2 \by = a \D_1\by + b \D_2\by + c \by ,
\end{equation}
where $\D_i = T_i - 1$, $i=1,2$, is the partial difference operator.
Intersections of tangent lines define two new quadrilateral surfaces called
the Laplace transforms of the old lattice.
\begin{Rem}
Restriction from $\PP^M$ to its affine part, and therefore from homogeneous
coordinates to non-homogeneous ones, results in putting $c=0$ in
equations~\eqref{eq:Lap}-\eqref{eq:D-Lap}.  
\end{Rem}

The tangents of the lattice are canonical examples of special two-parameter
families of straight lines called discrete congruences.
\begin{Def} \label{def:D-cong}
$\ZZ^2$-parameter family of lines in $\PP^M$ is called two dimensional
{\it discrete congruence} if any two neighbouring lines are coplanar.
\end{Def}
\begin{Rem}
Two neighbouring tangent lines $\langle [\by], T_i[\by] \rangle$ and
$T_j^{-1}\langle [\by], T_i[\by] \rangle$, $i\ne j$, of the quadrilateral
surface $[\by(m_1,m_2)]$ are coplanar and intersect giving 
$\cL_{ij}$ the Laplace transform of the lattice (see Fig.~\ref{fig:quadr2}).
\end{Rem}
\begin{Def}
Intersection points of lines of a discrete congruence with its nearest
neighbours in the $i$-th direction form the {\it $i$-th focal lattice} of 
the congruence. 
\end{Def}
One can show that focal lattices of two dimensional congruences are
quadrilateral lattices. The Laplace transformation can be considered as
correspondence between focal lattices of a congruence.

Similar notions and results exist in the continuous context.
\begin{Def} \label{def:cong}
Two-parameter family of lines in $\PP^M$ is called two dimensional
{\it congruence} if through each line pass two developable surfaces 
consisting of lines of the family.
\end{Def}
One can show that the curves of regression of such developables form two
{\it focal surfaces} tangent to the congruence, moreover the developables 
cut the focal surfaces along conjugate nets.

\section{Line geometry and the Pl\"ucker quadric}
\label{sec:line-geometry}
The interest of studying families of lines was motivated by the theory of
optics, and such mathematicians like Monge, Malus and Hamilton began to 
create
the general theory of rays. However it was Pl\"ucker, who first
considered straight lines in $\RR^3$ as primary elements; 
he also
found a convenient way to parametrize the
space of lines~\cite{Plucker}. The geometric interpretation of this 
parametrization 
was clarified later by Pl\"ucker's pupil Klein~\cite{Klein} and was one of
non-trivial examples in his Erlangen program. 

We present in this Section basic notions and
results of the line geometry, details can be found for example
in~\cite{Klein-VHG,Hlavaty}

The description of straight lines in $\RR^3$ takes more symmetric form
if we consider $\RR^3$ as the affine part of the projective space
$\PP^3$ (by the standard embedding $\by \mapsto [(\by,1)^T]$),
and study straight lines in that space. 
Given two different points $[\bu]$, $[\bv]$ of $\PP^3$, the line $\langle
[\bu], [\bv] \rangle$ passing
through them can be represented, up to proportionality factor, by a
bi-vector 
\begin{equation} \label{eq:bi-v-def}
\gp=\bu\wedge\bv \in \bigwedge^{2}(\RR^{4}) ;
\end{equation} 
changing the reference points of the line results in multiplying
the bi-vector by the determinant of the transition matrix between their
representatives. The space of straight lines in $\PP^3$ can be therefore
identified with a subset of $\PP\left(\bigwedge^{2}(\RR^{4})\right)
\simeq \PP^5$; the necessary and sufficient condition for a non-zero
bi-vector $\gp$ in order 
to represent a straight line is given by the homogeneous equation
\begin{equation} \label{eq:bi-v-simple}
\gp \wedge \gp = 0 ,
\end{equation}
a simple consequence of \eqref{eq:bi-v-def}.

If $\boe_1,\dots\boe_4$ is a basis of $\RR^4$ then the following
bi-vectors 
\begin{equation*} \boe_{i_1 i_2}=
\boe_{i_1}\wedge \boe_{i_2}, \qquad 1\leq i_1 < i_2\leq 4,
\end{equation*}
form the corresponding basis of $\bigwedge^{2}(\RR^{4})$:
\begin{equation*}
\gp = p^{12}\boe_{12} + p^{13}\boe_{13} + 
\dots + p^{34}\boe_{34} \; .
\end{equation*}
Equation~\eqref{eq:bi-v-simple} rewritten in the Pl\"ucker (or
Grassmann--Pl\"ucker) coordinates
$p^{ij}$ reads
\begin{equation} \label{eq:Pl-quad}
p^{12}p^{34}-p^{13}p^{24}+p^{14}p^{23} = 0 \; ,
\end{equation}
and defines in $\PP^5$ the so-called Pl\"{u}cker (or Pl\"ucker---Klein)
quadric $\cQ_P$. 

Let us present basic subsets of the quadric~$\cQ_P$
and corresponding configurations of lines in $\PP^3$.
\begin{enumerate}
\item If two lines intersect then the corresponding 
bi-vectors $\gp_i$, $i=1,2$, satisfy not only 
equations of the form~\eqref{eq:bi-v-simple}, 
but also
\begin{equation}
\gp_1\wedge \gp_2 = 0 ,
\end{equation} 
i.e., the corresponding points $[\gp_1]$, $[\gp_2]$
of the Pl\"{u}cker quadric are joined by an isotropic (i.e., contained in
$\cQ_P$) line.
Therefore isotropic lines of $\PP^5$ 
correspond to planar pencils of  lines in $\PP^3$.
\item A conic section of $\cQ_P$ by a non-isotropic plane represents the 
so called
{\it regulus}, i.e., one family of lines of a ruled quadric in $\PP^3$. 
\end{enumerate}

\section{Asymptotic nets and W-congruences in line geometry}
\label{sec:ass-nets}

We collect here, for Reader's convenience, 
various results of the theory of asymptotic 
nets~\cite{Bianchi,Eisenhart-TS,Lane,Finikov}, which we
consider necessary
to understand the methods and goals of next sections where we treat the
discrete case. 

\begin{Def}
A coordinate system on a surface in $\PP^3$ is called {\it asymptotic
parametrization} if in each point of the surface
the osculating planes of the parametric curves coincide
with the tangent plane to the surface.
\end{Def}
\begin{Rem}
Through the paper we consider asymptotic parametrization on a surface in the
projective space $\PP^3$, but we perform calculations in its affine part
$\RR^3$. 
\end{Rem}
Given a surface $\bx(u_1,u_2)$ in $\RR^3$ in asymptotic coordinates $u_1$,
$u_2$ then
\begin{align} \label{eq:as-1}
\pp_1^2 \bx& =  
a_1\pp_1 \bx + 
b_1\pp_2 \bx, \\
\label{eq:as-2}
\pp_2^2 \bx & = 
a_2 \pp_1 \bx + 
b_2\pp_2 \bx .
\end{align}
As a consequence of the compatibility condition 
$\pp_1^2\pp_2^2 \bx =
\pp_2^2\pp_1^2 \bx$ we obtain that there exists
a function $\phi(u_1,u_2)$ such that
\begin{equation*} \label{eq:a1b2-phi}
a_1 = \pp_1\phi, \qquad
b_2 = \pp_2\phi.
\end{equation*} 
The tangents to the asymptotic lines are represented, in the appropriate 
gauge, by the bi-vectors
\begin{equation*}
\gp_1 = e^{-\phi}\begin{pmatrix} \bx \\ 1 \end{pmatrix} \wedge 
\begin{pmatrix} \pp_1 \bx \\ 0 \end{pmatrix}, \qquad
\gp_2 = e^{-\phi}\begin{pmatrix} \bx \\ 1 \end{pmatrix} \wedge 
\begin{pmatrix} \pp_2 \bx \\ 0 \end{pmatrix};
\end{equation*}
notice that the line passing through $[\gp_1]$ and $[\gp_2]$ is an isotropic
line.

Equations \eqref{eq:as-1}-\eqref{eq:as-2} lead to the linear system
\begin{align} 
\label{eq:p11}
\pp_1 \gp_1 & = b_1\gp_2, \\
\label{eq:p22}
\pp_2 \gp_2 & = a_2\gp_1,
\end{align}
and, in consequence, to the Laplace equations
\begin{align*} 
\pp_1 \pp_2 \gp_1 & = \pp_2(\log b_1)\pp_1 \gp_1 + a_2 b_1 \gp_1, \\
\pp_1 \pp_2 \gp_2 & = \pp_1(\log a_2)\pp_2 \gp_2 + a_2 b_1 \gp_2 .
\end{align*}

The above results can be expressed as follows.
\begin{Th}
A surface in $\PP^3$ viewed as the envelope of its tangent planes
corresponds to a congruence of isotropic lines
of the Pl\"{u}cker quadric 
$\cQ_P$; the focal nets of the congruence represent asymptotic directions 
of the surface.
\end{Th}

Let us equip $\RR^3$ 
with the scalar product and consider the corresponding cross-product 
$\times$.
One can show that any asymptotic net $\bx(u_1,u_2)$ in $\RR^3$ 
can be considered as a solution of the linear system
\begin{align} \label{eq:x1N} 
\pp_1\bx = \pp_1 \bN \times \bN, \\
\label{eq:x2N} \pp_2\bx = \bN \times \pp_2 \bN,
\end{align}
where $\bN(u_1,u_2)$ is orthogonal to the surface and satisfies
equation
\begin{equation} \label{eq:N12qN}
\pp_1\pp_2 \bN = q\bN ,
\end{equation}
with a function $q(u_1,u_2)$.
Equation \eqref{eq:N12qN} was first studied by Moutard~\cite{Moutard}, and
equations \eqref{eq:x1N}-\eqref{eq:x2N} connecting solutions of 
the Moutard equation with asymptotic nets are known as the Lelieuvre
formulas~\cite{Lelieuvre}. 
\begin{Rem}
It should be mentioned that the Lelieuvre formulas can be settled down
within the pure affine (even projective) geometry without refering to
additional structures in the ambient space \cite{KoPin}
\end{Rem}
One can show that $\bN$ satisfies, in addition to the Moutard equation,
the following linear equations
\begin{align*} 
\pp_1^2\bN &= (\pp_1\phi) \pp_1 \bN - b_1 \pp_2 \bN + d_1 \bN, \\
\pp_2^2\bN &= -a_2 \pp_1 \bN + (\pp_2\phi) \pp_2 \bN + d_2 \bN,
\end{align*}
where
\begin{align*}
d_1& = \pp_2 b_1 + b_1 \pp_2\phi, \\
d_2 &= \pp_1 a_2 + a_2 \pp_1\phi,\\
\intertext{moreover}
q& = \pp_1\pp_2 \phi + b_1 a_2.
\end{align*}

Given scalar solution $\theta(u_1,u_2)$ of the Moutard
equations~\eqref{eq:N12qN}, consider the linear system
\begin{align} 
\label{eq:thN1}
\pp_1(\theta \widehat\bN) & = (\pp_1 \theta) \bN - \theta \pp_1\bN, \\
\label{eq:thN2}
\pp_2(\theta \widehat\bN) & = -(\pp_2 \theta) \bN + \theta \pp_2\bN,
\end{align}
compatible due to~\eqref{eq:N12qN}.
Cross-differentiation of equations \eqref{eq:thN1}-\eqref{eq:thN2} shows
that $\widehat\bN(u_1,u_2)$ satisfies another Moutard equation
\begin{equation}\label{eq:hN12qN}
\pp_1\pp_2 \widehat\bN = \widehat{q}\widehat\bN ,
\end{equation}
with the proportionality function $\widehat{q}(u_1,u_2)$ given by
\begin{equation*}
\widehat{q} = 
\frac{ \pp_1\pp_2 \widehat\theta  }{\widehat\theta }, 
\qquad \widehat\theta
= \frac{1}{\theta}.
\end{equation*}
The transition from $\bN$ to $\widehat\bN$ relating solutions of 
two Moutard
equations \eqref{eq:N12qN} and \eqref{eq:hN12qN} is called the Moutard 
transformation~\cite{Moutard}.

Simple calculation shows that the surface 
\begin{equation} \label{eq:hx}
\widehat\bx = \bx + \widehat\bN \times \bN,
\end{equation}
can be obtained from $\widehat{\bN}$ via the Lelieuvre formulas. 
Notice that the straight lines $\langle \bx, \widehat\bx \rangle$
are tangent to both surfaces in corresponding
points, i.e., the lines form the so called Weingarten (or W for short)
congruence. 

\begin{Def}
Two-parameter family of straight lines in $\PP^3$ tangent to two surfaces
in such a way that asymptotic coordinate lines on both surfaces correspond
is called {\it W--congruence}.
\end{Def}

There exists another way to find W-congruences tangent to a given
asymptotic net $\bx$.
Because $\theta\widehat\bN\times\bN$ is tangent to $\bx$ therefore
it can be decomposed as
\begin{equation*}
\theta\widehat\bN \times \bN = A \pp_1\bx + B \pp_2\bx ; 
\end{equation*}
the coefficients $A(u_1,u_2)$ and $B(u_1,u_2)$ of the above decomposition
define, together with $\bx(u_1,u_2)$, the W--congruence. It can be shown that
the coefficients satisfy the linear system
\begin{align}
\label{eq:A1B}
\pp_2 A & = -a_2 B ,\\
\label{eq:B2A}
\pp_1 B & = -b_1 A.
\end{align}

Finally, we consider W--congruences in the spirit of Pl\"{u}cker geometry.
The bi-vector
\begin{equation*}
\gq \propto \begin{pmatrix} \bx \\ 1 \end{pmatrix} \wedge
\begin{pmatrix} \widehat\bx \\ 1 \end{pmatrix}
\end{equation*}
represents W--congruence. The bi-vector $\gq$
in the gauge
\begin{equation*}
\gq = \theta e^{-\phi} \begin{pmatrix} \bx \\ 1 \end{pmatrix} \wedge
\begin{pmatrix} \widehat\bN\times\bN \\ 0 \end{pmatrix} = A\gp_1 + B\gp_2,
\end{equation*}
satisfies, due to linear systems \eqref{eq:p11}-\eqref{eq:p22} and
\eqref{eq:A1B}-\eqref{eq:B2A}, the Laplace equation
\begin{equation*}
\pp_1\pp_2\gq = (\pp_2\log B) \pp_1\gq + (\pp_1\log A) \pp_2 \gq +
\left[ a_2 b_1 - (\pp_1\log A) (\pp_2\log B) \right] \gq. 
\end{equation*}

\begin{Th} \label{th:W-cong}
W--congruences are represented by conjugate nets in the Pl\"ucker
quadric
$\cQ_P$.
\end{Th}

\section{Discrete asymptotic nets}
\label{sec:DA}
\begin{figure}
\begin{center}
\epsffile{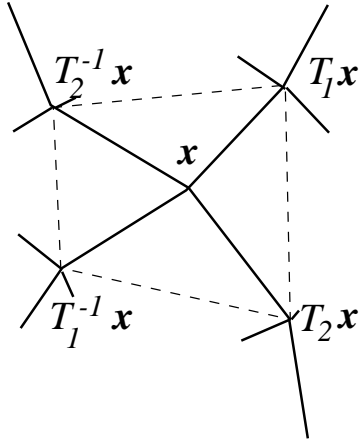}
\end{center}
\caption{Asymptotic lattice}
\label{fig:d-as}
\end{figure}

\begin{Def}[\cite{Sauer}] \label{def:d-as}
An {\it asymptotic lattice} is a mapping $\bx:\ZZ^2\to\RR^3$
such that any point $\bx$ of the lattice is coplanar with its four nearest 
neighbours $T_1\bx$, $T_2\bx$, $T_1^{-1}\bx$ and $T_2^{-1}\bx$ (see Fig.
\ref{fig:d-as}).
\end{Def}
The plane in Definition~\ref{def:d-as} can be called the
{\it tangent plane} of
the asymptotic lattice in the point $\bx$.  

\begin{figure}
\begin{center}
\epsffile{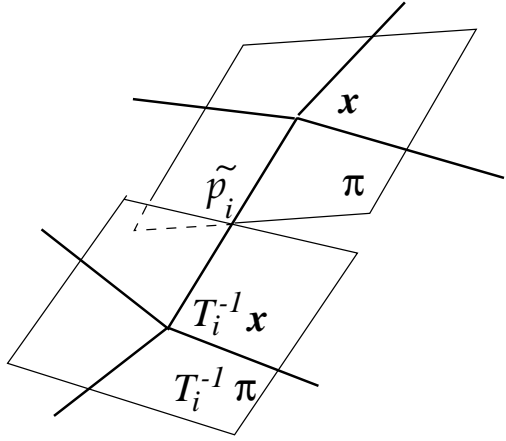}
\end{center}
\caption{Asymptotic directions as focal lattices of the isotropic congruence}
\label{fig:izot-cong}
\end{figure}

We can express the asymptotic lattice condition in the form of the linear
equations
\begin{align}
\label{eq:Dx11}
\D_1\tD_1\bx  &= a_1 \D_1\bx  +
b_1 \D_2\bx , \\
\label{eq:Dx22}
\D_2\tD_2\bx  &= a_2 \D_1\bx  +
b_2 \D_2\bx ,
\end{align}
where $\tD_i = 1 - T_i^{-1}$, $i=1,2$, is the backward partial
difference operator.
Equations~\eqref{eq:Dx11}-\eqref{eq:Dx11} 
can be rewritten using the backward tangent vectors as
\begin{align}
\label{eq:Dxt11}
\D_1\tD_1\bx &= \ta_1\tD_1\bx + \tb_1\tD_2\bx, \\
\label{eq:Dxt22}
\D_2\tD_2\bx &= \ta_2\tD_1\bx + \tb_2\tD_2\bx;
\end{align}
here the backward and forward data of the asymptotic lattice are related by
the following formulas
\begin{align*}
\ta_1 &= \frac{1-b_2}{D} -1, &
\tb_1 &= \frac{b_1}{D} ,\\
\ta_2 &= \frac{a_2}{D}, &
\tb_2 &= \frac{1-a_1}{D} -1,
\end{align*}
with
\begin{equation*}
D = (1-a_1)(1-b_2) - a_2 b_1 = \left((1+\ta_1)(1+\tb_2) - \ta_2 \tb_1
\right)^{-1} .
\end{equation*}
The compatibility condition of the linear
system~\eqref{eq:Dx11}-\eqref{eq:Dx22} leads, among others, to 
\begin{equation} \label{eq:Dx1122}
T_2^{-1}(1-a_1)T_2(1+\ta_1) = T_1^{-1}(1-b_2) T_1(1+\tb_2).
\end{equation}
The backward asymptotic tangent lines can be represented in the 
line geometry by the bi-vectors
\begin{equation*}
\tgp_i = \begin{pmatrix} \bx \\ 1 \end{pmatrix} \wedge
\begin{pmatrix} \tD_i\bx \\ 0 \end{pmatrix} , \qquad i=1,2.
\end{equation*}
Using equations~\eqref{eq:Dxt11}-\eqref{eq:Dxt22} it can be easily shown
that
\begin{align}
\label{eq:T1tp1}
T_1\tgp_1 & = (\ta_1 +1) \tgp_1 + \tb_1 \tgp_2, \\
\label{eq:T2tp2}
T_2\tgp_2 & = \ta_2 \tgp_1 + (\tb_2+1) \tgp_2.
\end{align}
Applying to equation~\eqref{eq:T1tp1} the shift operator $T_2$ and using
formulas \eqref{eq:T1tp1}-\eqref{eq:T2tp2} yields to the an equivalent form
of the discrete Laplace equation  
\begin{equation*}
T_1T_2\tgp_1 = (T_2\ta_1 +1)T_2\tgp_1 +
\frac{T_2\tb_1}{\tb_1}(\tb_2+1)T_1\tgp_1 - 
\frac{T_2\tb_1}{\tb_1 D}\tgp_1,
\end{equation*}
similarly we get
\begin{equation*}
T_1T_2\tgp_2 = (T_1\tb_2 +1)T_1\tgp_2 +
\frac{T_1\ta_2}{\ta_2}(\ta_1+1)T_2\tgp_2 - 
\frac{T_1\ta_2}{\ta_2 D}\tgp_2.
\end{equation*}
Notice that the lines  $\langle \tgp_1 , \tgp_2 \rangle$ are generators
of the Pl\"ucker quadric (both asymptotic tangents intersect in $\bx$) and
represent pairs $(\bx, \pi)$, where $\pi$ is the
tangent plane of the asymptotic lattice at the point $\bx$. Two
neighbouring tangent planes $\pi$ and $T_i^{-1}\pi$, $i=1,2$, intersect 
along the 
backward tangent line 
represented by $\tgp_i$ (see Fig.~\ref{fig:izot-cong}).
We have thus proved the following result:

\begin{Th} \label{th:D-A}
A discrete asymptotic net in $\PP^3$ viewed as the envelope of its tangent 
planes corresponds to a congruence of isotropic lines
 of the Pl\"{u}cker quadric 
$\cQ_P$; the focal lattices of the congruence represent asymptotic 
directions of the lattice.
\end{Th}
\begin{Cor}
The lattices in $\cQ_P$ which represent two families of asymptotic tangents 
of an asymptotic lattice are Laplace transforms of each other.
\end{Cor}

Similarly like in the continuous case there exist the discrete analogue 
of the Lelieuvre representation and the discrete analog of the Moutard 
equation; for details see~\cite{KoPin,NieszporskiDA}. It can be shown that
\begin{align}
\label{eq:D-L1}
\D_1\bx &= \D_1\bN\times\bN, \\
\label{eq:D-L2}
\D_2\bx &= \bN\times\D_2\bN,
\end{align}
where the vector $\bN$, orthogonal to the tangent plane of the lattice,
satisfies the discrete Moutard equation (see also~\cite{NiSchief})
\begin{equation*}
T_1T_2\bN + \bN = Q (T_1\bN + T_2\bN) ,
\end{equation*}
whose equivalent form is
\begin{equation}
\label{eq:DN12}
\D_1\D_2\bN = (Q-1) (\D_1\bN + \D_2\bN + 2\bN) .
\end{equation}
We would like to add some new ingredients to the connection of the Lelieuvre
representation of the asymptotic lattices and the linear
system~\eqref{eq:Dx11}-\eqref{eq:Dx22}. The normal vector satisfies
equations   
\begin{align}
\label{eq:DN11}
\D_1\tD_1\bN &= a_1\D_1\bN - b_1\D_2\bN + d_1\bN, \\
\label{eq:DN22}
\D_2\tD_2\bN &= -a_2\D_1\bN + b_2\D_2\bN + d_2\bN.
\end{align}
The compatibility condition of the system~\eqref{eq:DN11}-\eqref{eq:DN22}
with the Moutard equation~\eqref{eq:DN12} give
\begin{align}
\label{eq:Ttb-QQ}
(1-b_2) T_1(1+\tb_2)= & Q(T_2^{-1}Q) ,\\
\label{eq:Tta-QQ}
(1-a_1) T_2(1+\ta_1)= & Q(T_1^{-1}Q) .
\end{align}
Combining equations \eqref{eq:Ttb-QQ}-\eqref{eq:Tta-QQ} with 
\eqref{eq:Dx1122} yields the following identity
\begin{equation} \label{eq:Dx1122-n}
\frac{T_1 T_2\: \ta_1 + 1}{(T_1 Q)(T_2 D)(T_2\: \ta_1 + 1)} =
\frac{T_1 T_2\: \tb_2 + 1}{(T_2 Q)(T_1 D)(T_1\: \tb_2 + 1)} = F,
\end{equation}
which will be used in the next Section.

\section{Discrete W-congruences}
\label{sec:DW}
Similarly like in the continuous case, given solution $\Theta(n_1,n_2)$ 
of the discrete Moutard equation~\eqref{eq:DN12}, one can define 
the (discrete analog
of the) Moutard transformation~\cite{NieszporskiDA} (see
also~\cite{NiSchief}) by
solving the linear system
\begin{align}
\label{eq:tN-N1}
\D_1(\Theta \widehat\bN)& = (\D_1\Theta)\bN - \Theta \D_1\bN, \\
\label{eq:tN-N2}
\D_2(\Theta \widehat\bN)& = -(\D_2\Theta)\bN + \Theta \D_2\bN,
\end{align}
which yields
\begin{equation*}
T_1T_2\widehat\bN + \widehat\bN = \widehat{Q} (T_1\widehat\bN + 
T_2\widehat\bN), 
\end{equation*}
with new proportionality factor
\begin{equation*}
\widehat{Q}=\frac{T_1T_2\widehat\Theta + \widehat\Theta}{T_1\widehat\Theta +
 T_2\widehat\Theta}, \qquad \widehat\Theta = \frac{1}{\Theta}.
\end{equation*}
Let us define the following lattice 
\begin{equation} \label{eq:hx-NtN}
\widehat\bx = \bx + \widehat\bN\times\bN ;
\end{equation}
a simple calculation shows that formula~\eqref{eq:hx-NtN} gives 
new asymptotic lattice with 
the normal vector $\widehat\bN$ entering into the Lelieuvre formulas.

The line $\langle \bx, \widehat\bx \rangle$ is
tangent to both lattices, therefore we have
\begin{equation}
\label{eq:TtNxN}
\Theta\widehat\bN\times\bN = A\D_1\bx + B\D_2\bx = 
\tA\tD_1\bx + \tB\tD_2\bx,
\end{equation}
where
\begin{align}
\label{eq:tA-A}
\tA & = A(\ta_1 + 1) + B\ta_2,\\
\label{eq:tB-B}
\tB & = A\tb_1 + B(\tb_2 + 1). 
\end{align}
Notice that the two parameter family of lines 
$\langle \bx, \widehat\bx \rangle$ has analogous properties of that of
the {\it W--congruence} from continuous case. 
\begin{Def} \label{def:DW-cong}
By a {\it discrete W--congruence} we mean two-parameter family of straight
lines connecting two asymptotic lattices in such a way that the lines
are tangent to the lattices in corresponding points.
\end{Def}
We have shown how the get discrete W--congruences from the Moutard
transformations. It turns out that any discrete W--congruence can be 
obtained
in this way.
\begin{Prop} \label{prop:Wcong-Mout}
Given discrete W--congruence connecting $\bx$ and $\widehat{\bx}$,
then the normal vectors $\bN$ and $\widehat\bN$ which define $\bx$ and
$\widehat{\bx}$ via the Lelieuvre formulas, are related by a Moutard 
transformation.
\begin{equation*}
\begin{CD}
\bx @>{W-congruence}>> \widehat{\bx} \\
@VVV    @VVV  \\
\bN @>{\text{Moutard transf.}}>>  \widehat{\bN}
\end{CD}
\end{equation*}
\end{Prop}
\begin{proof}
From Definition \ref{def:DW-cong} it follows that $\widehat\bx$ must be of
the form
\begin{equation}
\label{eq:D-WM-p}
\widehat\bx = \bx + \psi \widehat\bN\times\bN.
\end{equation}
We first show that without loss of generality
one can put the proportionality function $\psi(n_1,n_2)$ equal to $1$.
Applying the partial difference operator
$\D_1$ to equation \eqref{eq:D-WM-p} and using the first part 
\eqref{eq:D-L1}
of the discrete Lelieuvre formulas we get
\begin{equation}
\label{eq:D-WM-p2}
T_1\widehat\bN \times \bN = T_1 \bN \times \bN + 
(T_1\psi)T_1\widehat\bN\times
T_1\bN - \psi \widehat\bN \times \bN.
\end{equation}
The scalar products with $T_1\widehat\bN$ and with $\bN$ give
\begin{align*}
(T_1 \bN \times \bN)\cdot T_1\widehat\bN &= \psi (\widehat\bN \times\bN) 
\cdot T_1\widehat\bN ,\\
(T_1\widehat\bN \times \widehat\bN)\cdot \bN &= T_1\psi (T_1 \bN \times \bN)
\cdot\bN,
\end{align*}
which after simple manipulation gives
\begin{equation}
\label{eq:psi1}
(T_1 \psi) \psi = 1;
\end{equation}
similarly we have
\begin{equation}
\label{eq:psi2}
(T_2 \psi) \psi = 1.
\end{equation}
Notice that due to equations \eqref{eq:psi1} and \eqref{eq:psi2} the
normal vector $\widehat\bN'=\psi\widehat\bN$ defines the same lattice
$\widehat\bx$, which shows that in formula \eqref{eq:D-WM-p} we can put
$\psi\equiv 1$.

After such a change, formula \eqref{eq:D-WM-p2} can be rewritten in the form
\begin{equation*}
(T_1\bN - \widehat \bN )\times (\bN - T_1\widehat \bN) = 0, 
\end{equation*}
which yields
\begin{equation}
\label{eq:D-WM-p3a}
T_1\bN - \widehat \bN = \lambda (\bN - T_1\widehat \bN).
\end{equation} 
Similarly we obtain
\begin{equation}
\label{eq:D-WM-p3b}
T_2\bN + \widehat \bN = \mu(\bN + T_2\widehat \bN).
\end{equation}
Formulas \eqref{eq:D-WM-p3a}-\eqref{eq:D-WM-p3b} give, together with the 
Moutard equations satisfied by $\bN$ and $\widehat\bN$, the following
equations
\begin{align*}
\lambda Q &= (T_2\lambda)\widehat{Q}, & \mu Q -1 &=(T_2\lambda)(\mu -
\widehat{Q} ),\\
\mu Q &= (T_1\mu)\widehat{Q}, & \lambda Q -1 &=(T_1\mu)(\lambda -
\widehat{Q} ) .
\end{align*}
This gives
\begin{equation*}
(T_2\lambda)\mu = (T_1\mu)\lambda ,
\end{equation*}
which implies that
\begin{equation} \label{eq:D-WM-p4}
\lambda = \frac{T_1\Theta}{\Theta} , \qquad \mu = \frac{T_2\Theta}{\Theta},
\end{equation}
moreover $\Theta$ satisfies the Moutard equation of $\bN$. Finally, equations
\eqref{eq:D-WM-p3a}-\eqref{eq:D-WM-p3a} with $\lambda$ and $\mu$ given by
\eqref{eq:D-WM-p4} can be put in the form of the Moutard
transformation~\eqref{eq:tN-N1}-\eqref{eq:tN-N2}.
\end{proof}
Equation~\eqref{eq:TtNxN} imply  
\begin{equation}
\label{eq:TtN}
\Theta\widehat\bN = \tA\tD_1\bN - \tB\tD_2\bN + \tC\bN.
\end{equation}
The compatibility condition of equation~\eqref{eq:TtN} and the Moutard
transformation~\eqref{eq:tN-N1}-\eqref{eq:tN-N2} gives, among others, 
\begin{align}
\label{eq:T1tB-B}
T_1\left( \frac{\tB}{\tb_2 + 1}\right) & = \frac{B}{Q} , \\
\label{eq:T2tA-A}
T_2\left( \frac{\tA}{\ta_1 + 1}\right) & = \frac{A}{Q}.
\end{align}
The following result is generalization of Theorem~\ref{th:W-cong}
to the discrete case.
\begin{Th} \label{th:D-W}
Discrete W--congruences are represented by two dimensional quadrilateral 
lattices in the Pl\"ucker quadric $\cQ_P$.
\end{Th}
\begin{proof} 
Lines of the W--congruence are represented by bi-vectors
\begin{equation*}
\gq = \begin{pmatrix} \bx \\ 1 \end{pmatrix} \wedge
\begin{pmatrix} \Theta \bN\times\bN \\ 0 \end{pmatrix} =
\tA\tgp_1 + \tB\tgp_2 .
\end{equation*}
We will show that $\gq$ satisfies the Laplace equation.

Because of~\eqref{eq:T1tp1}-\eqref{eq:T2tp2} we have
\begin{align*}
T_1\gq & = T_1 \tA \left[ (\ta_1 + 1) \tgp_1 + \tb_1 \tgp_2 \right] + 
(T_1\tB) T_1\tgp_2 ,\\
T_2\gq & = T_2 \tB \left[ (\tb_2 + 1) \tgp_2 + \ta_2 \tgp_1 \right] +
(T_2\tA) T_2\tgp_1
\end{align*}
and therefore
\begin{equation} \label{eq:q12}
T_1T_2\gq = \frac{T_1T_2 \tA}{T_2\tA}T_2(\ta_1 + 1) T_2\gq +
 \frac{T_1T_2 \tB}{T_1\tB}T_1(\tb_2 + 1) T_1\gq + U\tgp_1 + V\tgp_2,
\end{equation}
where
\begin{align*}
U &= 
\ta_2\frac{T_1T_2 \tA}{T_2\tA} T_2\left( \tA \tb_1 - 
\tB  (\ta_1 +1 )\right) + 
(\ta_1+1)\frac{T_1T_2 \tB}{T_2\tB} T_1\left( \tB \ta_2 - 
\tA (\tb_2 +1 )\right) \\
V &= 
(\tb_2+1)\frac{T_1T_2 \tA}{T_2\tA} T_2\left( \tA \tb_1 - 
\tB  (\ta_1 +1 )\right) + 
\tb_1\frac{T_1T_2 \tB}{T_2\tB} T_1\left( \tB \ta_2 - 
\tA (\tb_2 +1 )\right) .
\end{align*}
Using equations \eqref{eq:tA-A}-\eqref{eq:tB-B} we get
\begin{equation*}
U = 
-\ta_2\frac{T_1T_2 \tA}{T_2\tA} T_2\left( \frac{B}{D}\right) - 
(\ta_1+1)\frac{T_1T_2 \tB}{T_2\tB} T_1\left( \frac{A}{D}\right), 
\end{equation*}
which, due to 
equations~\eqref{eq:T1tB-B}-\eqref{eq:T2tA-A}, 
can be transformed to
\begin{equation*}
U = -Q (T_1 A)(T_2 B)\left( \frac{\ta_2 (T_1 T_2 \ta_1 + 1)}{A (T_1Q)(T_2 D)
(T_2\:\ta_1+1)} + \frac{(\ta_1 +1) (T_1 T_2 \tb_2 + 1)}{B (T_2Q)(T_1 D)
(T_1\:\tb_2+1)} \right). 
\end{equation*}
Identity \eqref{eq:Dx1122-n} gives, together with equations
\eqref{eq:tA-A}-\eqref{eq:tB-B}, that 
\begin{equation*}
U = -\tA Q F \frac{(T_1 A)(T_2 B)}{AB},
\end{equation*}
similarly,
\begin{equation*}
V = -\tB Q F \frac{(T_1 A)(T_2 B)}{AB},
\end{equation*}
which yields
\begin{equation} \label{eq:UV}
U\tgp_1 + V\tgp_2 =  - Q F \frac{(T_1 A)(T_2 B)}{AB} \gq .
\end{equation}
Inserting~\eqref{eq:UV} to equation~\eqref{eq:q12} leads to conclusion
that the bi-vector $\gq$ satisfies the Laplace equation.
\end{proof}
From interpretation of W--congruences as quadrilateral lattices in $\cQ_P$
we infer the following property (see final remarks of 
Section ~\ref{sec:line-geometry}).
\begin{Cor}
Four neighbouring lines of a W--congruence are generators of a ruled
quadric in $\PP^3$.
\end{Cor}

We would like to stress that the discrete
W--congruences are not discrete congruences in the sense of 
Definition~\ref{def:D-cong}.
In order to explain this terminological confusion 
we would like to make a few historical comments. At the
beginning of the line geometry, by a congruence it was meant any 
two-parameter
family of straight lines in $\RR^3$. It turns out that in $\RR^3$ such
family has, in general, two focal surfaces. However, in more dimensional 
ambient space two-parameter families of lines do not have, in general,
focal surfaces.
From the point of view of transformations of surfaces it was necessary, 
therefore, to put some
restrictions on the initial definition, and we 
read in~\cite[p. 11]{Eisenhart-TS}: "We
call a {\it congruence} in {\it n}-space a two-parameter family of lines 
such that through each line pass two developable surfaces of the family."
Going further into multi-parameter families of lines and into discrete
domain, in order to keep the basic property of congruences they 
have been defined~\cite{TQL} in such a way that they have focal 
lattices; this requirement leads to Definition~\ref{def:D-cong}. 
In continuous case W--congruences have focal surfaces, but this is, as we
mentioned above, typical property of two-parameter families of lines in
$\RR^3$. 
In our opinion, this terminological confusion suggests 
that it is more convenient to consider discrete
W--congruences as {\it quadrilateral lattices in the line space}.
 
\section{Permutability theorems} \label{sec:perm}

In this Section we consider superposition of the Moutard transformations
and the corresponding superpositions of
$W$-transformations of asymptotic lattices. 
We prove also the permutability theorems for both transformations. 

Let $\Theta^1(n_1,n_2)$ and $\Theta^2(n_1,n_2)$ be two solutions of the
Moutard equation of the lattice $\bN(n_1,n_2)$, i.e.,
\begin{equation} \label{eq:dM-1-2}
T_1T_2 \left( 
       \begin{array}{c} \bN \\ \Theta^1 \\ \Theta^2 \end{array}
       \right)  + \left( 
       \begin{array}{c} \bN \\ \Theta^1 \\ \Theta^2 \end{array}
       \right) = Q \left[ T_1\left( 
       \begin{array}{c} \bN \\ \Theta^1 \\ \Theta^2 \end{array}
       \right)  + T_2 \left( 
       \begin{array}{c} \bN \\ \Theta^1 \\ \Theta^2 \end{array}
       \right) \right] .
\end{equation}
We use $\Theta^1$ to define the first Moutard transformation $\bN_1$ of the
lattice $\bN$ and the corresponding transformation $\Theta^2_1$ of
$\Theta^2$ via equations \eqref{eq:tN-N1}-\eqref{eq:tN-N2}:
\begin{align} \label{eq:d1-N-N1}
\D_1\left[ \Theta^1  \left( \begin{array}{c}
\bN_1 \\ \Theta^2_1  \end{array} \right)  \right] & = 
(\D_1 \Theta^1)  \left( \begin{array}{c} \bN \\ 
\Theta^2 \end{array} \right)  -
\Theta^1 \D_1 \left( \begin{array}{c}  \bN \\ 
\Theta^2 \end{array} \right), \\
\label{eq:d2-N-N1}
\D_2\left[ \Theta^1  \left( \begin{array}{c}
\bN_1 \\ \Theta^2_1  \end{array} \right) \right] & = 
-(\D_2 \Theta^1) \left( \begin{array}{c}  \bN \\ 
\Theta^2 \end{array} \right) +
\Theta^1 \D_2  \left( \begin{array}{c} \bN \\ 
\Theta^2 \end{array} \right),
\end{align}
which implies that both $\bN_1$ and $\Theta^2_1$ satisfy the same Moutard
equation 
\begin{equation*}
T_1T_2 \left( 
       \begin{array}{c} \bN_1 \\  \Theta^2_1 \end{array}
       \right)  + \left( 
       \begin{array}{c} \bN_1  \\ \Theta^2_1 \end{array}
       \right) = Q_1 \left[ T_1\left( 
       \begin{array}{c} \bN_1  \\ \Theta^2_1 \end{array}
       \right)  + T_2 \left( 
       \begin{array}{c} \bN_1 \\ \Theta^2_1 \end{array}
       \right) \right] ,
\end{equation*}
where 
\begin{equation*}
Q_1=\frac{T_1T_2\widehat\Theta^1 + \widehat\Theta^1}{T_1\widehat\Theta^1 +
 T_2\widehat\Theta^1}, \qquad \widehat\Theta^1 = \frac{1}{\Theta^1}.
\end{equation*}
Similarly, we use $\Theta^2$ to define the second Moutard transformation 
$\bN_2$ of the
lattice $\bN$ and the corresponding transformation $\Theta^1_2$ of
$\Theta^1$:
\begin{align} \label{eq:d1-N-N2}
\D_1\left[ \Theta^2  \left( \begin{array}{c}
\bN_2 \\ \Theta^1_2  \end{array} \right)  \right] & = -
(\D_1 \Theta^2)  \left( \begin{array}{c} \bN \\ 
\Theta^1 \end{array} \right)  +
\Theta^2 \D_1 \left( \begin{array}{c}  \bN \\ \Theta^1 \end{array} \right), \\
\D_2\left[ \Theta^2  \left( \begin{array}{c}
\label{eq:d2-N-N2}
\bN_2 \\ \Theta^1_2  \end{array} \right) \right] & = 
(\D_2 \Theta^2) \left( \begin{array}{c}  \bN \\ 
\Theta^1 \end{array}\right) -
\Theta^2 \D_2  \left( \begin{array}{c} \bN \\ \Theta^1 \end{array} \right).
\end{align}
Notice the modification of signs in the transformation formulas, which however
do not change the fact that both $\bN_2$ and $\Theta^1_2$ satisfy the 
same Moutard equation 
\begin{equation*}
T_1T_2 \left( 
       \begin{array}{c} \bN_2 \\  \Theta^1_2 \end{array}
       \right)  + \left( 
       \begin{array}{c} \bN_2  \\ \Theta^1_2 \end{array}
       \right) = Q_2 \left[ T_1\left( 
       \begin{array}{c} \bN_2  \\ \Theta^1_2 \end{array}
       \right)  + T_2 \left( 
       \begin{array}{c} \bN_2 \\ \Theta^1_2 \end{array}
       \right) \right] ,
\end{equation*}
where 
\begin{equation*}
Q_2=\frac{T_1T_2\widehat\Theta^2 + \widehat\Theta^2}{T_1\widehat\Theta^2 +
 T_2\widehat\Theta^2}, \qquad \widehat\Theta^2 = \frac{1}{\Theta^2}.
\end{equation*}
Equations \eqref{eq:d1-N-N1}-\eqref{eq:d2-N-N2} imply that both products
$\Theta^1 \Theta^2_1$ and $\Theta^2 \Theta^1_2$ are defined up to additive
constants. Moreover, since
\begin{align*} 
\D_1(\Theta^1 \Theta^2_1) &= \D_1(\Theta^1) \Theta^2 - 
\Theta^1 \D_1\Theta^2 = \D_1(\Theta^2 \Theta^1_2) \\
\D_2(\Theta^1 \Theta^2_1) &= -\D_2(\Theta^1) \Theta^2 + 
\Theta^1 \D_2\Theta^2 = \D_2(\Theta^2 \Theta^1_2),
\end{align*}
then one of these constants can be fixed in such a way that
\begin{equation} \label{eq:pM-constr}
\Theta^1 \Theta^2_1= \Theta^2 \Theta^1_2 = \Xi^{12}
\end{equation}
holds.

The following result states that there exist 
lattices being simultaneous Moutard transformations of 
$\bN_1$ and $\bN_2$, what can be illustrated by the diagram
\begin{equation*}
\begin{CD}
\bN @>{\Theta^1}>> \bN_1 \\
@V{\Theta^2}VV    @VV{\Theta^2_1}V  \\
\bN_2 @>{\Theta^1_2}>>  \bN_{12}.
\end{CD}
\end{equation*} 
\begin{Th}[Permutability of the Moutard transformations] \label{th:perm-M}
Let $\Theta^1$, $\Theta^2$ be solutions of the discrete Moutard 
equation of the lattice $\bN$, and let $\bN_1$, $\bN_2$ be the corresponding
two (discrete) Moutard transformations of $\bN$. Then the functions 
$\Theta^1_2$
and $\Theta^2_1$, given by equations \eqref{eq:d1-N-N1}-\eqref{eq:pM-constr},
provide by the formula
\begin{equation} \label{eq:N12}
\bN_{12} + \bN = \frac{\Theta^1\Theta^2}{\Xi^{12}}
(\bN_1 + \bN_2) ,
\end{equation} 
one parameter family (because of the free integration constant
in $\Xi^{12}$) 
of the Moutard transformations of the lattice $\bN_1$ (by means of the function 
$\Theta^1_2$) which are
simultaneously the Moutard transformation of the 
lattice $\bN_2$ (by means of the function $\Theta^1_2$). 
\end{Th}
\begin{proof}
It is enough to verify directly that the lattice $\bN_{12}=\bN_{21}$ given
by \eqref{eq:N12}
is a solution of equations
\begin{align*}
\D_1(\Theta^1_2 \bN_{21})& = (\D_1\Theta^1_2)\bN_2 - \Theta^1_2 \D_1\bN_2, 
\\
\D_2(\Theta^1_2 \bN_{21})& = -(\D_2\Theta^1_2)\bN_2 + \Theta^1_2 \D_2\bN_2,
\end{align*}
which define the Moutard transformations of $\bN_2$ by means of
$\Theta^1_2$, and that it is also a solution of equations
\begin{align*}
\D_1(\Theta^2_1 \bN_{12})& = -(\D_1\Theta^2_1)\bN_1 + \Theta^2_1 \D_1\bN_1, 
\\
\D_2(\Theta^2_1 \bN_{12})& = (\D_2\Theta^2_1)\bN_1 - \Theta^2_1 \D_2\bN_1,
\end{align*}
which define the Moutard transformations of $\bN_1$ by means of
$\Theta^2_1$.
\end{proof}
\begin{Rem}
Notice that the superposition formula \eqref{eq:N12} itself is of the
form of the discrete Moutard equation. This is a manifestation of the 
frequently observed relation between discrete integrable systems and their
Darboux-type transformations
\cite{LeBen,NiSchief,BP1,GanzhaTsarev,KoSchief2,TQL}. To obtain such a form
of the superposition formula it was the reason of modification of signs in
the Moutard transformation \eqref{eq:d1-N-N2}-\eqref{eq:d2-N-N2}.  
\end{Rem}
The corresponding theorem (the discrete analog of the classical Bianchi
permutability theorem \cite{Bianchi,Eisenhart-TS})
about permutability of the $W$-transformations of
asymptotic lattices reads as follows.
\begin{Th}[Permutability of the $W$-transformations] \label{th:perm-W}
If $\bx$ and $\bx_1$ are asymptotic lattices related by a $W$-congruence and
$\bx$ and $\bx_2$ are related by a second $W$-congruence then there can be
found one-parameter family of asymptotic lattices given, in notation of
Theorem \ref{th:perm-M}, by 
\begin{equation} \label{eq:x_12}
\bx_{12}=\bx_{21} = \bx + \frac{\Theta^1\Theta^2}{\Xi^{12}}
\bN_1\times\bN_2,
\end{equation}
such that $\bx_1$ and $\bx_{12}$ are related by a $W$-congruence, and
likewise $\bx_2$ and $\bx_{12}$.
\end{Th}
\begin{proof}
Due to Proposition \ref{prop:Wcong-Mout} the lattices $\bx_1$ and $\bx_2$
can be given by
\begin{align} \label{eq:W1-N1}
\bx_1 & = \bx + \bN_1\times\bN , \\
\label{eq:W2-N2}
\bx_2 & = \bx - \bN_2\times\bN ,
\end{align}
where $\bN$, $\bN_1$ and $\bN_2$ are like in
\eqref{eq:d1-N-N1}-\eqref{eq:d2-N-N2}; notice the change of sign in 
\eqref{eq:W2-N2} induced by the change of sign in 
\eqref{eq:d1-N-N2}-\eqref{eq:d2-N-N2}. Transforming lattice $\bx_1$ 
by \eqref{eq:W2-N2} by means of $\bN_{12}$ and applying \eqref{eq:N12} one
obtains \eqref{eq:x_12}, likewise transforming lattice $\bx_2$ 
by formula \eqref{eq:W1-N1} by means of $\bN_{12}$.
\end{proof}

\section{Conclusion and remarks}
\label{sec:concl}
The main result of this paper consists in showing that the theory of
asymptotic lattices and their transformations given by W--congruences
forms a part of the theory of quadrilateral lattices. The 
discrete W--congruences can be considered as
quadrilateral lattices in the Pl\"ucker quadric, therefore they provide
non-trivial examples of quadrilateral lattices subjected to quadratic
constraints, whose general theory was constructed in~\cite{q-red}.
We demonstrated also the permutability property of the corresponding
$W$-transformations of asymptotic lattices, thus proving directly their
integrability.

Our result is the next step in realization of the general program of 
classification
of integrable geometries as reductions of quadrilateral lattices. Such
reductions come usually from additional structures in the projective
ambient space (the close analogy to the Erlangen program of Klein), and/or 
from inner symmetries of the lattice itself (see also examples in
\cite{DS-sym}); in our case the basic underlying geometry is
the line geometry of Pl\"ucker. 

\section*{Acknowledgments}
The author wishes to express his thanks to Maciej Nieszporski for 
discussions
and for showing preliminary version of his paper~\cite{NieszporskiDA}.

\bibliographystyle{amsplain}
\providecommand{\bysame}{\leavevmode\hbox to3em{\hrulefill}\thinspace}

\end{document}